\documentclass{aa}
\usepackage{amssymb}

\usepackage{graphicx}

\begin{document}

\title{Limits on diffusive shock acceleration in supernova remnants in the presence
of cosmic-ray streaming instability and wave dissipation}
\author{V.S. Ptuskin \and V.N. Zirakashvili}
\date{Received ; accepted }

\titlerunning{Limits on diffusive shock acceleration in supernova remnants}
\subtitle{}


\offprints{V.S. Ptuskin, \\ e-mail: vptuskin@izmiran.rssi.ru}

\institute{Institute for Terrestrial Magnetism, Ionosphere
   and Radiowave Propagation of the Russian Academy of Sciences, Troitsk,
   Moscow Region 142190, Russia}

\abstract{ The instability in the cosmic-ray precursor of a supernova shock
moving in interstellar medium is studied. The level of magnetohydrodynamic
turbulence in this region determines the maximum energy of particles
accelerated by the diffusive shock acceleration mechanism. The high efficiency
of cosmic ray acceleration is accepted, and the consideration is not limited by
the case of weak turbulence. It is assumed that Kolmogorov type nonlinear wave
interactions together with the ion-neutral collisions restrict the amplitude
of random magnetic field. As a result, the maximum energy of accelerated
particles strongly depends on the age of supernova remnant. The maximum energy can be as high
as $\sim10^{17}Z$ eV in young supernova remnants and falls down to about
$\sim10^{10}Z$ eV at the end of Sedov stage. Thus the standard estimate of
maximum particle energy based on the Bohm limit calculated for the
interstellar magnetic field strength is not justified in this case. This
finding may explain why the supernova remnants with the age more than a few
thousand years are not prominent sources of very high energy gamma-rays.

   \keywords{supernova remnants -- cosmic rays -- acceleration of particles --
                shock waves -- turbulence -- gamma rays}
   }

%
\maketitle 
\section{Introduction}

Particle acceleration by supernova blast waves is commonly accepted as
the main mechanism of acceleration of galactic cosmic rays, see Drury et al.
(\cite{drury01}), Malkov \& Drury (\cite{malkov01}) for review. The high density of energetic
particles in the shock vicinity where diffusive shock acceleration
occurs leads to the modification of gas flow through the action of energetic
particle pressure and initiates plasma instabilities produced by the current
of energetic particles. The plasma instabilities tend to increase the level
of magnetohydrodynamic turbulence which scatter the particles. In the case
of strong turbulence $\delta B\sim B_{0}$ ($\delta\mathbf{B}$ is the
amplitude of random magnetic field, $\mathbf{B}_{0}$ is the regular magnetic
field upstream of the shock), it may decrease the cosmic ray diffusion
coefficient $D$ to the Bohm value $D_{B}=vr_{g}/3$, where $v$ is the particle
velocity, $r_{g}=pc/(ZeB_{0})$ is the Larmor radius of a particle with
momentum $p$ and charge $Ze$. The Bohm value of diffusion coefficient is
commonly used in simulations of cosmic ray acceleration in supernova
remnants, e.g. Berezhko et al. (\cite{berezhko96}).

The dependence of diffusion on energy determines the maximum energy which
particles can gain in the process of acceleration. The necessary condition
of efficient acceleration at the shock at the free expansion stage is
$D(E)\lesssim0.1u_{sh}R_{sh}$,
where $R_{sh}$ is the radius and $u_{sh}=dR/dt$
is the velocity of spherical shock. Notice that the typical supernova burst
with kinetic energy of ejecta $\mathcal{E=E}_{51}10^{51}$ erg in the
interstellar gas with number density $n_{0}$ cm$^{{-3}}$ gives the maximum
value of the product $0.1u_{sh}R_{sh}\sim10^{27}(\mathcal{E}
_{51}/n_{0})^{0.4}$ cm$^{\mathrm{2}}$ s$^{-1}$ at the end
of the free expansion stage of
supernova remnant evolution when this product reaches its maximum value. At
the same time, the typical value of cosmic ray diffusion coefficient in the
Galaxy is
$D_{ISM}=5.9\times10^{28}\beta R_{m}^{0.3}$ cm$^{\mathrm{2}}$ s$^{-1}$,
where $\beta=v/c$ is the dimensionless particle velocity and $R_{m}=pc/Z$
is the particle magnetic rigidity in units GV, see
Jones et al. (\cite{jones01}), and thus the necessary condition of acceleration can
not be fulfilled for relativistic particles unless their diffusion
coefficient is anomalously small in the vicinity of the shock. The Bohm
value which is a lower bound of diffusion coefficient along the magnetic
field at $\delta B\lesssim B_{0}$ is equal to $D_{B}=6\times10^{21}\beta
R_{m}$ cm$^{\mathrm{2}}$/s at $B_{0}=5\times10^{-6}$ G. The entire pattern of cosmic ray acceleration
on supernova shocks critically depends on the assumption that the energetic
particles themselves produce the turbulence needed to provide the
anomalously slow diffusion at the cite of acceleration. This was perceived at
the very beginning of the studies of diffusive shock acceleration mechanism
(Bell, \cite{bell78}, Lagage \& Cesarsky, \cite{lagage83}).

Analyzing the early stage of supernova remnant evolution when the shock
velocity is high, $u_{sh}\thicksim10^{4}$ km s$^{-1}$, Lucek \& Bell
(\cite{lucek00}), Bell \& Lucek (\cite{bell01})
found that the cosmic ray streaming instability can be so
strong that the amplified field $\delta B\gtrsim10^{-4}$ G far exceeds the
magnetic field in the undisturbed medium ahead of the moving shock
$B_{0}\sim (2\div5)\times10^{-6}$ G
(see also earlier papers of McKenzie \&
V\"{o}lk (\cite{mckenzie82}), V\"{o}lk et al. (\cite{voelk84}),
Falle \& Giddings (\cite{falle87}), Bennett
\& Ellison (\cite{bennett95})).
The maximum particle energy increases accordingly. The
cosmic ray acceleration continues at later stage of supernova evolution when
the mass of swept-up interstellar gas exceeds the mass of supernova ejecta
and the shock velocity goes down according to Sedov solution. The major
portion of galactic cosmic rays is accelerated by the expanding blast wave
during this stage. The cosmic-ray streaming instability is less
efficient as the shock velocity falls down with time and the nonlinear
wave interactions can significantly reduce the level of turbulence at late
Sedov stage (V\"{o}lk et al. \cite{voelk88},
Fedorenko \cite{fedorenko90}). This leads to fast
diffusion and decreases the maximum energy which can be achieved in a
process of diffusive shock acceleration. This effect is aggravated by the
wave damping which is due to the ion-neutral collisions if neutral atoms are
present in the interstellar gas upstream of the shock
(Bell, \cite{bell78}; Drury et
al. \cite{drury96}).

In the present paper, we consider the acceleration of cosmic rays and their
streaming instability in a wide range of shock velocities. The level of
magnetic field fluctuations is allowed to be arbitrary large ($\delta
B\gtrless B_{0}$), and the rate of nonlinear wave interactions is assumed to
correspond the Kolmogorov type nonlinearity. The dissipation of waves by the
ion-neutral collisions is also taken into account in the case of supernova
burst in not fully ionized interstellar gas. The task is to find the maximum
energy of accelerated particles as a function of age of supernova remnant.
The problem is enormously complicated by the fact that the processes of
wave-particle and wave-wave interactions have no universally accepted theoretical
description in the case of strong turbulence. So, our consideration is made
at the level of approximate estimates and we present not more than a
schematic description of the overall picture.

\section{Maximum energy of accelerated particles}

We consider the adiabatic stage of supernova remnant expansion when the
major fraction of cosmic rays injected in the interstellar medium is
accelerated. The evolution of a shock produced by supernova
outburst in interstellar medium is described by Sedov solution, see e.g.
Lozinskaya (\cite{lozinskaya92}), Ostriker \& McKee
(\cite{ostriker98}). In the case of very efficient
acceleration when the cosmic ray pressure inside the supernova envelope
governs its expansion, the equations for $R_{sh}(t)$ and $u_{sh}(t)$ are:
\begin{eqnarray}
R_{sh} &=&4.26\left( \frac{\mathcal{E}_{51}}{n_{0}}\right) ^{1/5}\left(
\frac{t}{10^{3}\mathrm{ yr}}\right) ^{2/5}\mathrm{ pc,}\;  \nonumber \\
u_{sh} &=&1.67\times 10^{3}\left( \frac{\mathcal{E}_{51}}{n_{0}}\right)
^{1/5}\left( \frac{t}{10^{3}\mathrm{ yr}}\right) ^{-3/5}\mathrm{ km\ s^{-1}}.
\end{eqnarray}

The selfsimilar solution (1) is valid under the condition that the mass of
swept-up gas significantly exceeds the mass of ejecta $M_{ej}$ $(\sim
1M_{\odot})$ that is fulfilled at $R_{sh}>R_{0}=(3M_{ej}/4\pi
m_{a}n_{0})^{1/3}=1.9n_{0}^{-1/3}$ pc , $t>t_{0}=R/u_{0}=190n_{0}^{-1/3}$
yr, where $n_{0}$ is the number density of hydrogen atoms in the
interstellar gas, $m_{a}=1.4m$ is the mean interstellar atom mass per
hydrogen nucleus, $m$ is the proton mass, and $u_{0}\sim10^{4}$ km s$^{-1}$ is the
ejecta initial velocity. The preceding evolution of a supernova remnant
can be approximated as a free expansion with velocity $u_{0}$. The adiabatic
approximation used in the derivation of equations (1) is not valid when the
shock velocity sinks below about $u_{sh}=280\mathcal{E}
_{51}^{0.055}n^{0.111} $ km s$^{-1}$ and the radiation energy losses become
essential.{\Large \ }The evolution of a supernova shell with intense
radiative cooling goes as $R_{sh}\varpropto t^{2/7}$, $u_{sh}\varpropto
t^{-5/7}$. We set $\mathcal{E}_{51}=1$ in the subsequent numerical estimates.

In the following, we distinguish $n_{0}$, the mean gas number density which
determines the shock wave expansion law, and $n$, the intercloud gas number
density. It corresponds to cloudy structure of interstellar medium with
small dense clouds randomly distributed in the diffuse background gas. Two
phases of interstellar gas are considered below: a hot ionized component with the
temperature of background gas $T=7\times10^{5}$K, the number densities $%
n_{0}=10^{-2}$ cm$^{\mathrm{-3}}$, $n=3\times10^{-3}$ cm$^{\mathrm{-3}}$, and the
magnetic field strength $B_{0}=2$ $\mu$G; and a warm partly ionized component with the following
parameters: $T=8\times10^{3}$K, $n_{0}=0.4$ cm$^{\mathrm{-3}}$, $n=0.1$ cm$^{%
\mathrm{-3}}$, $n_{i}=0.03$ cm$^{\mathrm{-3}}$, $B_{0}=5$ $\mu$G (where $n_{i}$
is the number density of ions). We assume that the diffusive shock
acceleration of energetic particles should cease to operate after
$10^{5}$
years in the hot gas when the shock Mach number falls below $M\sim 3$,
and by the time $5\times10^{4}$ yr in the warm gas when
Sedov stage ends.

The acceleration of a fast particle diffusing near the shock front is a
version of Fermi type acceleration. The fast particles are scattered on
inhomogeneities of magnetic field frozen into background plasma and gain
energy crossing the shock where plasma is compressing. In the test particle
approximation, the distribution of accelerated cosmic ray particles on
momentum for high Mach number shocks has
the canonical power-law form $f(p)\varpropto p^{-4}$
up to some maximum momentum $p_{\max }$. In the
case of efficient acceleration, the action of cosmic ray pressure on the
shock structure causes nonlinear modification of the shock that changes the
shape of particle spectrum making it flatter at relativistic energies, see
Berezhko \& Ellison (1999). With allowance made for the nonlinear
modification, we assume in the following that the spectrum of energetic
particles at the shock is of the form
$f(p)\varpropto p^{-4+a}$ where $
0<a<0.5$. The normalization of function $f(p)$ is such that the integral
$N=4\pi \int dpp^{2}f(p)$ gives the number density of relativistic
particles. We assume that the pressure of cosmic rays at the shock $P=(4\pi
/3)\int dpp^{3}vf(p)$ is some fraction $\xi _{cr}\leq 1$ of the upstream
momentum flux entering the shock front, so that $P=\xi _{cr}\rho u_{sh}^{2}$
($\rho =m_{a}n_{0}$). The typical value of $\xi _{cr}\thickapprox 0.5$ and
the total compression ratio $\sim 7$ were found in the numerical simulations
of strongly modified shocks by Berezhko et al. (1996) under the assumption
of efficient dissipation of Alfv\'enic turbulence in the shock precursor. Thus
we accept the following distribution function of cosmic rays at the shock:
\begin{eqnarray}
f_{0}(p) &=&\frac{3\xi _{cr}\rho u_{sh}^{2}H(p_{\max }-p)}{4\pi c\varphi
(p_{\max })}\frac{1}{(mc)^{a}p^{4-a}},\mathrm{\ }  \nonumber \\
\varphi (p_{\max }) &=&\int\limits_{0}^{p_{\max }/mc}\frac{dyy^{a}}{\sqrt{%
1+y^{2}}},
\end{eqnarray}
where $H(p)$ is the step function. The ultra relativistic asymptotics of $%
\varphi (p)$ is $\varphi (p)\approx (p/mc)^{a}/a$ and the good approximation
at $p/mc\gtrsim 1$ is $\varphi (p_{\max })\approx a^{-1}(p_{\max
}/mc)^{a}(1-((1+a)(p/mc)^{a})^{-1})$. The value $a=0.3$ is used in the numerical estimates below.

The spatial distribution of accelerated particles in the plane shock
approximation can be roughly presented as
\begin{equation}
f_{1}(p,x)=f_{0}(p)\exp \left( u_{sh}\int\limits_{0}^{x}\frac{dx_{1}}{%
D(p,x_{1})}\right)
\end{equation}
at $x\leq 0$ (upstream of the shock); and
\begin{equation}
f_{2}(p,x)=f_{0}(p)
\end{equation}
at $x\geq 0$ (downstream of the shock).

Here we use the reference frame where the shock is at rest at $x=0$, the
gas velocity is parallel to the $x$ axis, and $u_{x}>0$. The steady state
one-dimensional solution (3) is valid for $\left| x\right| \lesssim
\min\{R_{sh},$ $R_{sh}\left( D/uR\right) ^{1/2}\}$. Equation (3) shows the
presence of cosmic ray precursor of characteristic size $D/u_{sh}$
ahead of the shock.

The cosmic ray diffusion coefficient is determined by the following
approximate equation which can be used for an arbitrary strong turbulence, $%
\delta B\lessgtr B_{0}$:
\begin{equation}
D=\frac{\left( 1+A_{tot}^{2}\right) ^{1/2}}{3A^{2}(>k_{res})}%
vr_{g},\;k_{res}=\left( 1+A_{tot}^{2}\right) ^{1/2}r_{g}^{-1}.
\end{equation}
We introduce here the dimensionless amplitude
of the random field
$A_{tot}^{2}=\delta B^{2}/B_{0}^{2}=
\left( 4\pi /B_{0}^{2}\right) \int dkW(k)$
, where $W(k)$ is the energy density of magnetohydrodynamic turbulence. The
Larmor radius is defined through the regular field $r_{g}=pc/ZeB_{0}$. The
scattering of particles of Larmor radius $r_{g}$ is mainly produced by
random inhomogeneities with the resonant wave number $k_{res}$ and the
amplitude of resonant waves is characterized by $A^{2}(>k_{res})=4\pi
k_{res}W(k_{res})/B_{0}^{2}$. We set up the formula (5) with the rounded
numerical coefficient as the generalization of equation for cosmic ray
diffusion coefficient in weak random field (Berezinskii et al.,
\cite{berezinskii90}). In doing
the generalization, we distinguish for each energetic particle with momentum $p$ a large scale magnetic field,
 which includes the regular magnetic field plus the averaged random magnetic field with $k<k_{res}(p)$,
 and a small scale magnetic field, which includes the random magnetic field with $k>k_{res}(p)$. The
particle moves adiabatically in a large scale field and experiences scattering by small scale field.
 Note that the choice of numerical coefficient in (5) affects the numerical
coefficient in the equation (7) below.

It is usually assumed that the random field which is necessary for cosmic
ray diffusion inside the supernova shell can be produced by the advection of
turbulence through the shock front, by the intrinsic instabilities at the
shock transition region, and by the irregular gas flow downstream of the
shock. The turbulence in the upstream region should be due to the cosmic ray
streaming instability.

In the case under study, the following steady-state equation determines the
energy density $W(k,x)$ of the turbulence amplified by the streaming instability
in cosmic ray precursor upstream of the shock:
\begin{equation}
u\frac{\partial W}{\partial x}=2(\Gamma _{cr}-\Gamma _{l}-\Gamma _{nl})W.
\end{equation}
$x\leq 0$.

Here the left hand side of the equation describes the advection of
turbulence by gas flow with highly supersonic velocity $u=u_{sh}$. The terms
on the right hand side of the equation describe respectively the wave
amplification by cosmic rays, the linear damping of waves in background
plasma, and the nonlinear wave-wave interactions which limit the amplitude
of turbulence. In the strict sense, $\Gamma_{nl}$ is some
integral-differential operator acting on the function $W(k,x)$.

The diffusion current which is due to the nonuniform distribution of cosmic
ray particles upstream of the shock causes the streaming instability and the
amplification of magnetohydrodynamic waves with the growth rate
\begin{eqnarray}
\Gamma _{cr}(k) &=&\frac{12\pi ^{2}Z^{2}e^{2}V_{a}\left(
1+A_{tot}^{2}\right) ^{1/2}}{c^{2}k}  \nonumber \\
&&\int\limits_{p_{res}(k)}^{\infty }dpp\left( 1-\left( \frac{p_{res}(k)}{p}%
\right) ^{2}\right) D\left| \frac{\partial f}{\partial x}\right|
\end{eqnarray}
(the waves with $k_{x}<0$ are amplified). Here $p_{res}(k)=ZeB_{0}\left(
1+A_{tot}^{2}\right) ^{1/2}/ck$, and the Alfv\'en velocity is defined through
the regular field $B_{0}$ as $V_{a}=B_{0}/\sqrt{4\pi \rho }=1.8(B_{0}/1\mu $G%
$)n^{-1/2}$ km s$^{-1}$. The most abundant proton component of cosmic rays
determines the value of $\Gamma _{cr}$, so we put $Z=1$ in the subsequent
calculations. The expression (7) is the generalization of equation for weak
random field (Berezinskii et al, 1990) to the case of strong random field.

Equations (2), (3), (5), (7) let us estimate the value of growth rate in the
upstream region $(x\leq 0)$ as
\begin{eqnarray}
\Gamma _{cr}(k) &\approx &\frac{C_{cr}(a)\xi _{cr}u_{sh}^{3}k^{1-a}\left(
1+A_{tot}^{2}\right) ^{-(1-a)/2}}{cV_{a}\varphi (p_{\max })r_{g0}^{a}}
\nonumber \\
&&\exp \left( 3u_{sh}\int\limits_{0}^{x}dx_{1}\frac{kA^{2}(>k,x_{1})}{%
v\left( 1+A_{tot}^{2}\right) }\right) .
\end{eqnarray}
at $k\geq k_{res}(p_{\max })$, and $\Gamma _{cr}=0$ otherwise. Here $%
r_{g0}=mc^{2}/(eB_{0})$, $C_{cr}(a)=\frac{27}{4(5-a)(2-a)}$, so that $%
C_{cr}(0.3)\approx 0.845$.

Considering that all waves upstream of the shock are generated by cosmic-ray
streaming instability, the particles with maximum momentum $p_{\max}$ are in
resonance with the waves which have the minimum wave number $k_{\min}=\left(
1+A_{tot}^{2}\right) ^{1/2}r_{g}^{-1}(p_{\max})$. Note that $A^{2}(>k_{\min
})=A_{tot}^{2}$.

The most important mechanism of linear damping of Alfv\'enic turbulence in the
problem under consideration is the wave dissipation due to ion-neutral
collisions, see e.g. Kulsrud \& Cesarsky (1971). It occurs in not fully
ionized gas and is described by the equations
\begin{equation}
\Gamma _{l}=\frac{\nu _{in}}{2},k>\nu _{in}(1+n_{i}/n_{H})/\left( \sqrt{%
1+A_{tot}^{2}}V_{a}\right) ,\;
\end{equation}
\[
\Gamma _{l}=\frac{k^{2}(1+A_{tot}^{2})V_{a}^{2}}{2\nu
_{in}(1+n_{i}/n_{H})^{2}},k\ll \frac {\nu _{in}(1+n_{i}/n_{H})}
{\sqrt{1+A_{tot}^{2}}V_{a}} ,
\]
where $\nu _{in}=n_{H}\left\langle v_{th}\sigma \right\rangle \thickapprox
8.4\times 10^{-9}\left( T/10^{4}\mathrm{K}\right) ^{0.4}(n_{H}/1$ cm$^{-3})$ s$%
^{\mathrm{-1}}$ for temperatures $T\sim 10^{2}$K to $10^{5}$K is the frequency
of ion-neutral collisions with the cross section $\sigma $ averaged over
velocity distribution of thermal particles, $n_{H}$ is the number density of
neutral hydrogen. This gives $\nu _{in}=$ $7.7\times 10^{-10}$ s$^{\mathrm{-1}%
} $ in warm diffuse gas.

In spite of a great progress in the investigation of magnetohydrodynamic
turbulence, it is not easy to specify $\Gamma _{nl}$ in equation (6). The
incompressible MHD simulations of driven and decaying turbulence by Verma et
al. (\cite{verma96}) demonstrated the presence of Kolmogorov type energy cascade to large
wave numbers that can be qualitatively described by the equation
$$
\Gamma _{nl}W^{\pm }(k)\thickapprox
$$
\begin{equation}
-\frac{1}{2}C_{K}^{-3/2}\frac{\partial }{
\partial k}\left( k^{2}V_{a}\left( \frac{kW^{\mp }(k)}{B_{0}^{2}/4\pi }
\right) ^{1/2}W^{\pm }(k)\right) ,
\end{equation}
where $C_{K}\approx 3.6$ is the so-called ''Kolmogorov constant'' and $\pm $
signs correspond to Alfv\'en waves travelling in opposite directions, i.e.
with $k_{x}>0$ and $k_{x}<0$ respectively, the total wave energy density is $%
W(k)=W^{+}(k)+W^{-}(k)$, and the turbulence is assumed to be weak, $kW^{\mp }(k)\left(
B_{0}^{2}/4\pi \right) ^{-1}\ll 1$. The nonlinear cascade of Alfv\'enic waves
is anisotropic. The main part of energy density in this turbulence is
concentrated at perpendicular to the local magnetic field wave vectors $%
k_{\perp }\approx k$, while the parallel wave numbers are small:
$
k_{\parallel }\sim \left( kW(k)\left( B_{0}^{2}/4\pi \right) ^{-1}\right)
^{1/2}k_{\perp }$, see
Goldreich \& Sridhar (\cite{goldreich95}),
Cho \& Vishniak (\cite{cho00}).

The theoretical approach based on the kinetic theory of weakly turbulent
collisionless plasma gives somewhat different results, see Livshits \&
Tsytovich (\cite{livshits70}),
Lee \& V\"{o}lk (\cite{lee73}), Kulsrud (\cite{kulsrud78}),
Achterberg (\cite{achterberg81}), Fedorenko et al.
(\cite{fedorenkoetal90}), Zirakashvili (\cite{zirakashvili00}).
In particular, it was
shown that the nonlinear interactions of Alfv\'en waves and thermal
particles
lead to the nonlinear wave damping which is described by the equation
\begin{equation}
\Gamma _{nl}W^{\pm }(k)=1.1\sqrt{\beta _{th}}kV_{a}\left( \frac{kW^{\mp }(k)%
}{B_{0}^{2}/4\pi }\right) W^{\pm }(k)
\end{equation}
in the case of high $\beta _{th}=v_{T}^{2}/V_{a}^{2}>1$ plasma
(Zirakashvili, \cite{zirakashvili00})
and by the same within an order of magnitude expression
but without factor $\sqrt{\beta _{th}}$ in the case of low
$\beta _{th}\ll 1$
(Livshits \& Tsytovich, \cite{livshits70};
Zirakashvili, \cite{zirakashvili00}). Equation (11) was
obtained for the case of weak turbulence and under the assumption that wave
distribution is not strongly anisotropic in $k$-space. A rate similar to (11)
was derived for the wave-wave interactions in magnetohydrodynamics based
on the theory of perturbations at small wave amplitude
(Chin \& Wentzel, \cite{chin72};
Akhiezer et al., \cite{akhieser75}; Skilling, \cite{skilling75}).

The observations show that the turbulence in the solar wind
plasma is of
Kolmogorov type, that evidently favors equation (10)
to equation (11), and
contains Alfv\'en waves moving both along the magnetic field and in
perpendicular directions (Saur \& Bieber, \cite{saur99}).
The data on interstellar
turbulence are also consistent with the assumption that a single
Kolmogorov spectrum extends from scales $10^{8}$ to $10^{20}$ cm
(Armstrong et al., \cite{armstrong95}).

It is clear from expressions (10) and (11) that Alfven waves with the opposite
signs of $k_{x}$ are needed for nonlinear interactions. (The only exception
is the one-dimensional case when the interactions are allowed for the waves
propagating in one direction along the external magnetic field, see
Lee \& V\"{o}lk (\cite{lee73}), Kulsrud (\cite{kulsrud78}),
Achterberg (\cite{achterberg81}) and
Zirakashvili (\cite{zirakashvili00})
for corresponding equations and discussion.) The cosmic-ray streaming
upstream of the shock amplifies waves $W^{-}$ with $k_{x}<0$ that propagate
in the direction of cosmic ray diffusion flux, and damps the waves $W^{+}$
with $k_{x}>0$. So, one needs an additional mechanism to maintain the waves
in both directions and to assure their efficient nonlinear interactions.
This is provided by the scattering of waves on the fluctuations of gas
density. Considering the case of the most strong wave interactions, we
assume in the following that the isotropization of waves is fast and the
waves moving in both directions along the external magnetic field are
present. A similar approach is used in the studies of Alfv\'en waves in the
solar wind turbulence, see Goldstein et al. (\cite{goldstein95}),
Hu et al., (\cite{hu99}).

The Kolmogorov type nonlinearity is assumed in the consideration that
follows. The simplified expression

\begin{equation}
\Gamma _{nl}=(2C_{K})^{-3/2}V_{a}kA(>k)\approx 0.05V_{a}kA(>k),
\end{equation}
at $C_{K}=3.6$ is accepted in our calculations. The order of magnitude
estimate (12) is based on equation (10) and is valid for an arbitrary strong
turbulence.

It is worth noting that the earlier consideration of cosmic ray streaming
instability in supernova remnants by Lagage \& Cesarsky
(\cite{lagage83}), V\"{o}lk et
al. (\cite{voelk88}) and Fedorenko (\cite{fedorenko90}) was made in the limit of $\delta B<B_{0}$ and
with the use of equation similar to (11) that is very different from our
present approach.

Let us consider separately the cases of nonlinear and linear dissipation of
waves.

{\it A. Cosmic-ray streaming instability with Kolmogorov type
nonlinearity. }

The linear damping is ignored in this case. The wave density is determined
by equation (6) at $\Gamma _{l}=0$ and with $\Gamma _{nl}$ given by equation
(12). The exponential distribution of cosmic rays (3) with a
characteristic scale $D/u_{sh}$ determines the spatial distribution of waves
in the upstream region that gives the estimate of the first term of equation
(6) at the shock $u\partial W/\partial x\approx u^{2}W/D$ in the case of
efficient wave amplification. Now using equations (5), (8), and (12), one
can find from (6) the following approximate equation that determines the
amplitude of random field at the shock:
\begin{eqnarray}
\frac{3u_{sh}^{2}A^{2}(>k)}{2v(1+A_{tot}^{2})}+(2C_{K})^{-3/2}V_{a}A( &>&k)=
\nonumber \\
\frac{C_{cr}(a)\xi _{cr}u_{sh}^{3}}{cV_{a}\varphi (p_{\max
})(kr_{g0})^{a}\left( 1+A_{tot}^{2}\right) ^{(1-a)/2}}.
\end{eqnarray}

Equation (13) allows to find the following equation for $A_{tot}$:
\begin{eqnarray}
\frac{3u_{sh}^{2}A_{tot}^{2}}{2v(1+A_{tot}^{2})}+(2C_{K})^{-3/2}V_{a}A_{tot}
&=&  \nonumber \\
\frac{C_{cr}(a)\xi _{cr}u_{sh}^{3}}{cV_{a}\varphi (p_{\max })(p_{\max
}/mc)^{-a}\left( 1+A_{tot}^{2}\right) ^{1/2}}.
\end{eqnarray}

As was pointed out in the Introduction, the efficient particle acceleration
at the shock necessitates sufficiently small diffusion coefficient. The
maximum momentum of accelerated particles at the Sedov stage of SNR
evolution is approximately determined by the condition $D(p_{\max
})=\varkappa u_{sh}R_{sh}$, where $\varkappa \approx 0.04$, see e.g.
Berezhko et al. (\cite{berezhko96}). The particles with larger diffusion coefficient are
not accelerated and not confined in the remnant. The last condition and
equation (5) give
\begin{equation}
\frac{p_{\max }}{mc}=\frac{3\varkappa A_{tot}^{2}u_{sh}R_{sh}}{\left(
1+A_{tot}^{2}\right) ^{1/2}vr_{g0}}.
\end{equation}

As the most abundant species, protons mainly drive the instability. Other
particles experience resonant scattering on produced turbulence and undergo
diffusive shock acceleration. As a result, the ions with charge $Z$ have the
same shape of spectra as protons but written as function of $p/Z$ instead of
$p$. This is of course valid for the maximum momentum attained at the shock
and $p_{\max}/Z$ should appear in equation (15) instead of $p_{\max}$ if
$Z\neq1$ (as before,
$m$ is the proton mass, $r_{g0}$ is the Larmor radius of
a proton with momentum $mc$ in equation (15)). The very high energy
electrons are accelerated with the same rate as protons at the same magnetic
rigidity. However, the maximum energy of electrons is smaller then for
protons because of heavy energy losses on synchrotron radiation and
inverse Compton scattering, see analysis of these processes by Gaisser et
al. (\cite{gaisser98}), Reynolds (\cite{reynolds98}).

Equations (14), (15) together with equations (1) determine the maximum
momentum of accelerated particles (protons) during Sedov stage of SNR
evolution in an implicit form. The results of numerical solution of these
equation for $p_{\max}$ as the function of shock velocity $u_{sh}$ are
presented in Figures 1 and 2 for the cases of supernova remnant expansion in
hot and warm interstellar gas respectively.
The values of $\xi_{cr}=0.5$
, $a=0.3$, and $\mathcal{E}_{51}=1$ were accepted in these calculations. For
comparison, the solution without wave damping,
$\Gamma_{l}=0$, $\Gamma_{nl}=0 $ is presented.
The Bohm limit for maximum momentum calculated
at the interstellar magnetic field strength $B_{0}$ is also shown
in Figures 1, 2. The Bohm limit
represents the standard estimate of maximum
momentum of accelerated particles that is commonly accepted in the
investigations of cosmic ray acceleration in supernova remnants. Our
consideration shows that this ''standard '' estimate of $p_{\max}$ might be
completely unsuitable. It does not work in young supernova remnants at the
age $t\lesssim3\times10^{3}$ yr when cosmic ray stream instability generates
strong random magnetic field $\delta B>B_{0}$ that reduces particle
diffusion (in agreement with Bell \& Lucek (2001) results). It also does
not work in old supernova remnants at $t\gtrsim3\times10^{3}$ yr when strong
Kolmogorov type nonlinear interactions of waves damp the turbulence
generated by instability and thus increases particle diffusion upstream of
the shock.

It is useful to give simple analytic expressions for $p_{\max}$ in the
limits of high and low values of shock velocity $u_{sh}$.

It is easy to check that if the shock velocity is high enough so that
\begin{equation}
u_{sh} \gg \frac{4aC_{cr}(a)\xi _{cr}c}{9\left( 2C_{K}\right) ^{3/2}},
u_{sh} \gg \frac{3V_{a}}{2aC_{cr}(a)\xi _{cr}}
\end{equation}
then the first term in the left-hand side of equation (14) dominates over
the second term, and the wave amplitude is large, $A_{t}>>1$. Thus the
advection of waves is more important than the nonlinear wave interactions
and the turbulence is strong when (16) is fulfilled. The inequalities (16)
are satisfied at $u_{sh}\gg 8.3\times 10^{7}$ cm s$^{-1}$,
$u_{sh}\gg 7.9\times
10^{7}$ cm s$^{-1}$ in hot gas, and $u_{sh}\gg 8.3\times 10^{7}$ cm s$^{-1}$,
$u_{sh}\gg
3.4\times 10^{7}$ cm s$^{-1}$ in warm gas at $a=0.3$, $\xi _{cr}=0.5.$ The
equations (14), (15) now give
\begin{eqnarray}
\frac{p_{\max }}{mc} &\approx &\frac{2\varkappa aC_{cr}(a)\xi
_{cr}u_{sh}^{2}R_{sh}}{r_{g0}V_{a}c}\approx  \nonumber \\
&&2.2\times 10^{5}\left( \xi _{cr}/0.5\right) \mathcal{E}
_{51}^{3/5}n_{0}^{-3/5}n^{1/2}t_{\mathrm{Kyr}}^{-4/5}.
\end{eqnarray}
Here we set $\varkappa =0.04$, $a=0.3$, $C_{K}=3.6$, use equations (1) for $%
u_{sh}(t)$ and $R_{sh}(t)$, and approximate $\varphi (p_{\max })\approx
(p_{\max }/mc)^{a}/a$, $v=c;$ $t_{\mathrm{Kyr}}$ is the supernova remnant age
in units $10^{3}$ yr. Equation (17) estimates $p_{\max }/mc\approx 2\times
10^{5}t_{\mathrm{Kyr}}^{-4/5}$ in hot gas, and $p_{\max }/mc\approx 1\times
10^{5}t_{\mathrm{Kyr}}^{-4/5}$ in warm gas at $t_{\mathrm{Kyr}}\ll 3$, $\mathcal{%
E}_{51}=1$,$\xi _{cr}=0.5$. The cosmic ray diffusion coefficient in the
upstream region depends on rigidity as $D\varpropto r_{g}^{1-a}$ at
$p\lesssim p_{\max }$.

In the opposite limit of low shock velocity
\begin{eqnarray}
u_{sh} &\ll &\left( \frac{4V_{a}^{3}c^{2}}{\pi aC_{cr}(a)\left(
2C_{K}\right) ^{3}\xi _{cr}}\right) ^{1/5},\;  \nonumber \\
u_{sh} &\ll &\left( \frac{V_{a}^{2}c}{aC_{cr}(a)\left( 2C_{K}\right)
^{3/2}\xi _{cr}}\right) ^{1/3}
\end{eqnarray}
the second term in the left-hand side of equation (14) dominates over the
first term and the wave amplitude is small, $A_{t}\ll 1$. The non-linear
wave interactions are more important than advection in this case. The
inequalities (18) are satisfied at $u_{sh}\ll 7.4\times 10^{7}$ cm s$^{-1}$,
$u_{sh}\ll 8.0\times 10^{7}$ cm s$^{-1}$
in hot gas, and $u_{sh}\ll 4.4\times 10^{7}$
cm s$^{-1}$, $u_{sh}\ll 4.5\times 10^{7}$ cm s$^{-1}$ in warm gas.
The equations (14), (15) give
\begin{eqnarray}
\frac{p_{\max }}{mc} \approx \frac{0.24\varkappa
a^{2}C_{cr}^{2}(a)C_{K}^{3}\xi _{cr}^{2}u_{sh}^{7}R_{sh}}{
r_{g0}V_{a}^{4}c^{3}}\approx  \nonumber \\
4.1\times 10^{11}\left( \xi _{cr}/0.5\right) ^{2}\mathcal{E}
_{51}^{8/5}n_{0}^{-8/5}n^{2}B_{0,\mu \mathrm{G}}^{-3}t_{\mathrm{Kyr}}^{-19/5}
\end{eqnarray}
that leads to the estimate $p_{\max }/mc\approx 7\times 10^{8}
t_{\mathrm{Kyr}
}^{-19/5}$ in hot gas, and $p_{\max }/mc\approx 1
\times 10^{8}t_{\mathrm{Kyr}
}^{-19/5}$ in warm gas at $t_{\mathrm{Kyr}}\gg 10.$ The cosmic ray diffusion
coefficient in the upstream region depends on rigidity as $D\varpropto
vr_{g}^{1-2a}$ at $p\lesssim p_{\max }$ in this case.

\begin{figure*}[tbp]
\centering
\includegraphics[width=10.0cm,angle=270]{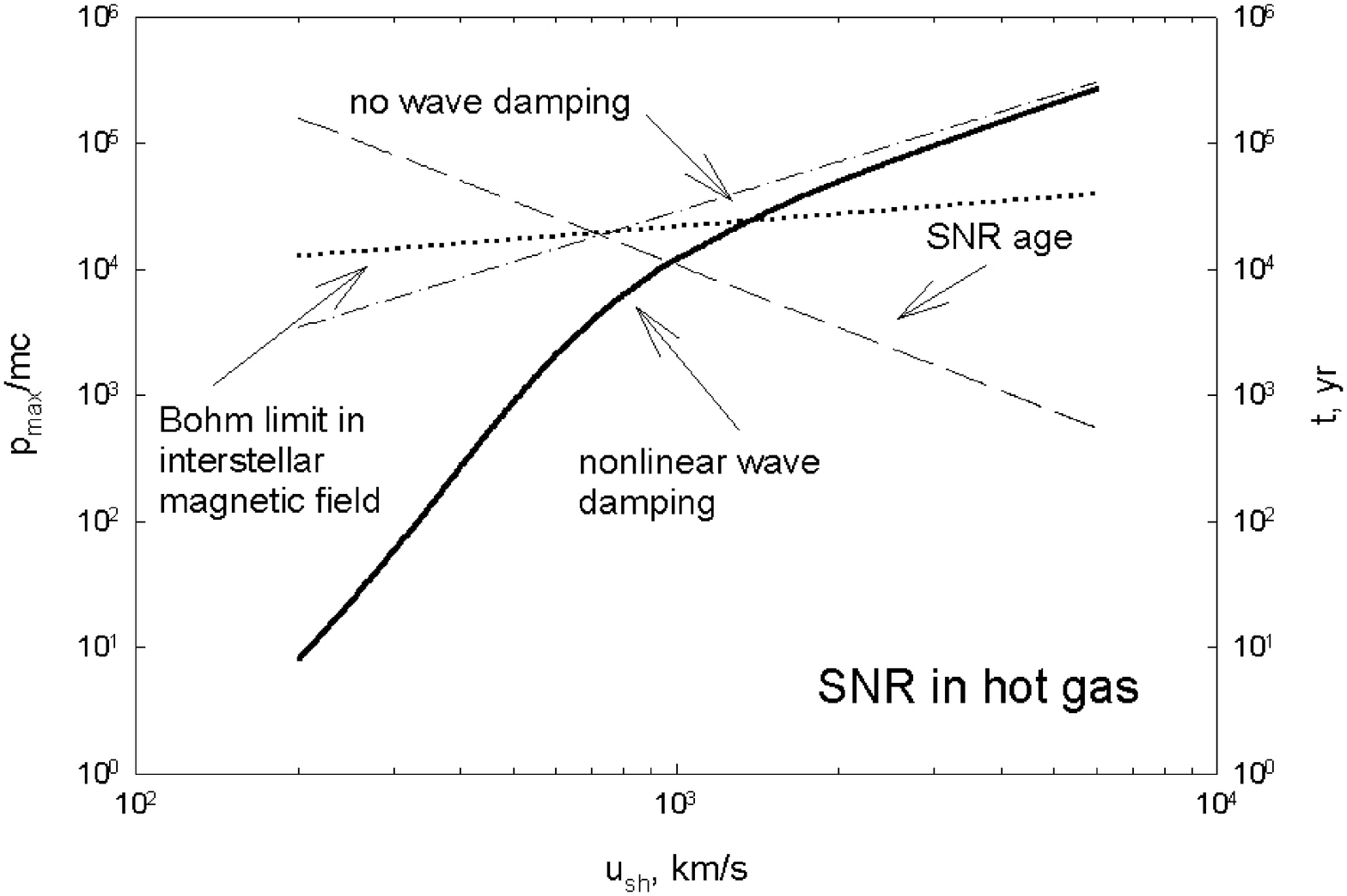}
\caption{The maximum momentum of accelerated protons $p_{\max}$ in units of $mc$ as
a function of shock velocity $u_{sh}$ at the Sedov stage of supernova
remnant evolution in the hot interstellar gas is shown by solid line. The
kinetic energy of the explosion is $\mathcal{E}=10^{51}$ erg; the efficiency
of cosmic ray acceleration $\protect\xi_{cr}=0.5$. The parameters of
interstellar medium are the following: the gas temperature $T=7\times10^{5}$
K, the average gas density $n_{0}=10^{-2}$ cm$^{-3}$, the density of diffuse
gas $n=3\times10^{-3}$ cm$^{-3}$, and the interstellar magnetic field strength $%
B_{0}=2$ $\protect\mu$G. The dash line presents the age of a supernova
remnant $t$ (ploted on the right ordinate) as a function of shock velocity.
Also shown: dot line - the Bohm limit on maximum particle momentum in the
field $B_{0}$; dash-dot line - the maximum particle momentum when the wave
dissipation is not taken into account. }
\label{Fig1}
\end{figure*}

As it is evident from Figures 1, 2 and equations (17), (19), the value of $%
p_{\max}$ is rapidly decreasing with the age of supernova remnant $t$. The
maximum energy of cosmic rays accelerated by supernova shock at early
Sedov stage is close to $3\times10^{14}$ eV for protons and it
falls down to about $10^{10}$ eV at the end of Sedov stage for the set of
parameters accepted in the present work. In particular, the particle energy
is less than $10^{13}$ eV for $t\gtrsim5\times10^{3}$ yr and this may
explain the absence of TeV gamma-ray signal from many supernova remnants
(Buckley et al. \cite{buckley98})
where gamma-rays could be produced through $\pi^{0}$
decays if sufficiently energetic protons and nuclei were present. The
commonly accepted estimate of maximum particle energy based on the Bohm
diffusion limit enormously overestimates $p_{\max}$ for these objects if the
level of self-generated turbulence is limited by the strong Kolmogorov type
nonlinearity (12).

\begin{figure*}[tbp]
\centering
\includegraphics[width=10.0cm,angle=270]{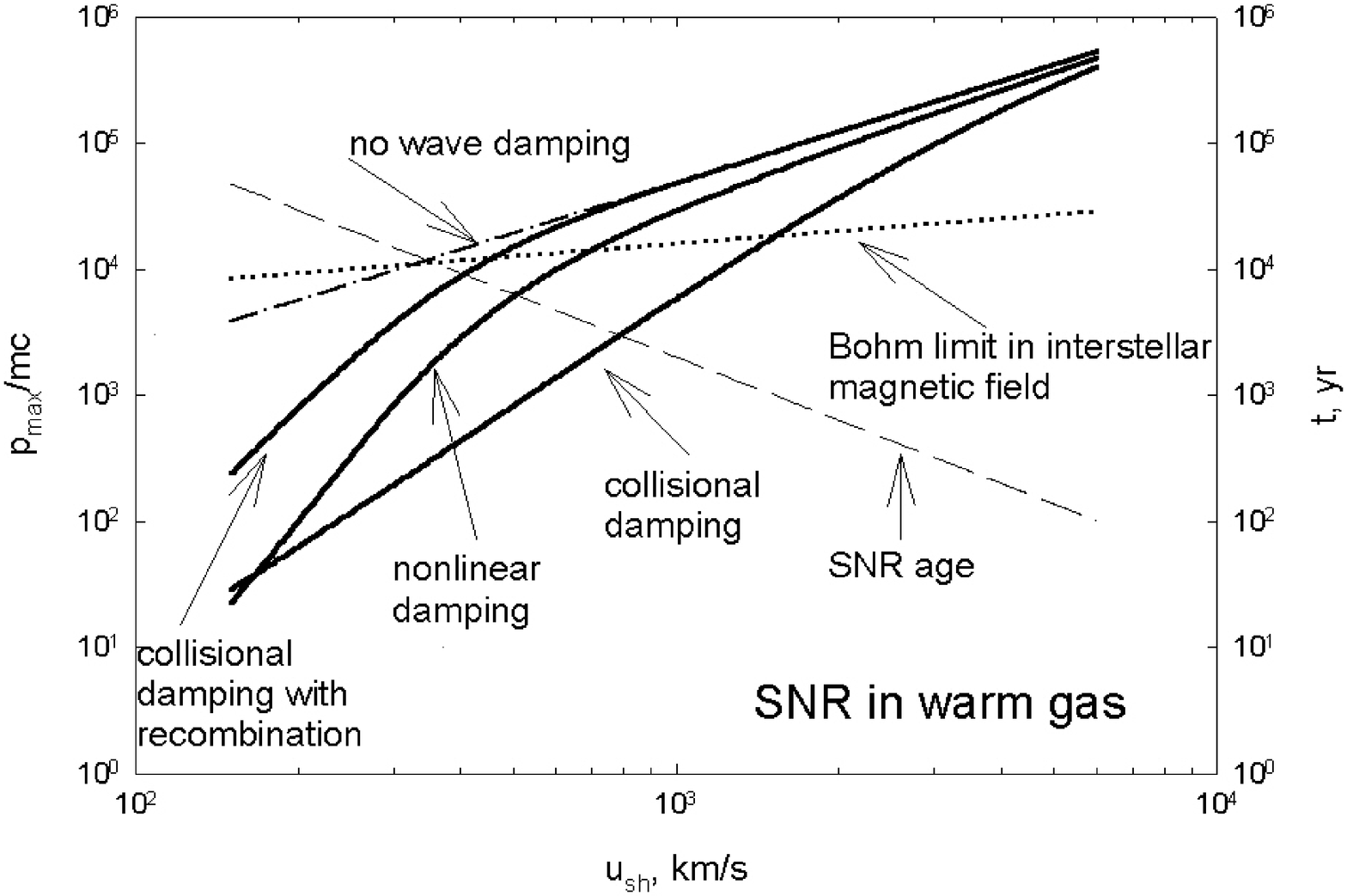}
\caption{The same as in Figure 1 but for the supernova burst in warm
interstellar gas ($T=8\times10^{3}$ K, $n_{0}=0.4$ cm$^{-3}$, $n=0.1$ cm$
^{-3}$, $B_{0}=5$ $\protect\mu$G). Three solid lines correspond to three
cases of wave dissipation considered separately: nonlinear wave
interactions; damping by ion-neutral collisions at given above number
density of neutral atoms; damping by ion-neutral collisions when the diffuse
neutral gas restors its' density after complete ionization by the radiation
from supernova burst.}
\label{Fig2}
\end{figure*}

{\it B. Cosmic-ray streaming instability in warm gas with collisional
wave dissipation.}

The cosmic ray acceleration in supernova remnants in the presence of wave
dissipation by ion-neutral collisions was investigated by Drury et al.
(1996). Different aspects of particle acceleration by supernova shocks
moving in partly ionized interstellar gas were also considered
by Boulares
\& Cox (\cite{boulares88}),
Bykov et al. (\cite{bykov00}). It was assumed that the turbulence
generated by cosmic ray streaming instability is weak. Here we do not restrict
our consideration to the case of weak turbulence.
We ignore the nonlinear wave dissipation in this section and consider the
collisional dissipation in partly ionized warm gas.
The wave density $W(k)$
is determined now from equation (6) where the term $\Gamma_{nl}$ is omitted
and the term $\Gamma_{l}$ is given by equation (9). So, similarly to the
case {\it A} and associated equations (13), (14), we have the following
equation for the amplitude of random magnetic field at the shock:

\begin{eqnarray}
\frac{3u_{sh}^{2}A^{2}(>k)k}{2v(1+A_{tot}^{2})}=  \nonumber \\
\frac{C_{cr}(a)\xi _{cr}u_{sh}^{3}k}{cV_{a}\varphi (p_{\max
})(kr_{g0})^{a}\left( 1+A_{tot}^{2}\right) ^{(1-a)/2}}-\frac{\nu _{in}}{2}
\end{eqnarray}
at $k\geq k_{\ast }$, $k_{\ast }=\nu _{in}/\left( \sqrt{1+A_{tot}^{2}}%
V_{a}\right) $.

Now, using equations (5), (15), (20) and relations $A^{2}(>k_{\min
})=A_{tot}^{2}$, $k_{\min }=\sqrt{1+A_{tot}^{2}}/r_{g,\max }$ we find the
following equation for $p_{\max }$:
\begin{equation}
\left( \frac{p_{\max }}{mc}\right) ^{1-a}\varphi (p_{\max })=\frac{
2C_{cr}(a)\xi _{cr}u_{sh}^{3}}{cV_{a}r_{g0}\left( \nu _{in}+\frac{u_{sh}}
{\varkappa R_{sh}}\right) },
\end{equation}
valid for $p_{\max }\leq p_{\ast }$, $p_{\ast }/mc=\left(
1+A_{tot}^{2}\right) V_{a}\left( \nu _{in}r_{g0}\right) ^{-1}$. The term
$\nu _{in}$ should be omitted
in equation (21) if $p_{\max }>p_{\ast }$. The
dissipation on neutrals is not essential for the development of instability
if the last inequality is fulfilled.

The value of maximum particle momentum $p_{\max }$ as a function of the
supernova shock velocity $u_{sh}$ determined according to equation (21) is
shown in Figure 2. The approximate expression based on equation (21) at $\nu
_{in}>20u_{sh}/R_{sh}$ is
\begin{eqnarray}
\frac{p_{\max }}{mc} \approx \frac{2aC_{cr}(a)\xi _{cr}u_{sh}^{3}}{
cV_{a}r_{g0}\nu _{in}}\approx  \nonumber \\
3\times 10^{3}\left( \xi _{cr}/0.5\right) \mathcal{E}
_{51}^{3/5}n_{0}^{-3/5}n^{-1/2}t_{\mathrm{Kyr}}^{-9/5}
\end{eqnarray}
at $a=0.3$ that gives $\left( p_{\max }/mc\right) \approx 5.7\times
10^{3}\left( u_{sh}/10^{3}\mathrm{ km\ s^{-1}}\right) ^{3}\approx 1.64\times
10^{4}t_{\mathrm{Kyr}}^{-9/5}$ at $\xi _{cr}=0.5$, $\mathcal{E}_{51}=1$. The
last approximate formula and Figure 2 evidence that the maximum energy of
accelerated particles decreases from about $3\times 10^{14}$ eV in early
Sedov stage to $3\times 10^{10}$ eV at the end of Sedov stage due to the
damping of cosmic-ray generated turbulence by ion-neutral collisions.

The solution with a smaller value of $p_{\max}$ at given shock velocity
$u_{sh}$ should be chosen when both nonlinear and collisional wave damping
are present.

It is known that some envelopes of Type Ia supernovae expand in partly
ionized interstellar gas (see Lozinskaya, 1992), so that the equations
obtained above are appropriate in this instance. On the other hand, an
explosion of Type II supernova leads to the full ionization of surrounding
diffuse gas over distances of several of tens parsecs that holds about
$10^{5}n^{-1}$ years after the burst. Therefore, the
intercloud neutral gas density is very low shortly after the burst and rises
linearly in time because of recombination as
\begin{equation}
n_{H}=C_{r}n^{2}t,
\end{equation}
at $t\ll \left( C_{r}n\right) ^{-1}$, where $C_{r}(T)$ is the recombination
rate of hydrogen. At a later time $t\thicksim \left( C_{r}n\right) ^{-1}$
the value of $n_{H}$ tends to some constant value which is determined by the
balance between ionization and recombination. The standard value of
recombination rate $C_{r}=4.1\times 10^{-13}$ cm$^{\mathrm{-3}}$s$^{\mathrm{-1}}$
at $T\approx 10^{4}K$ (Seaton, \cite{seaton59})
gives the time-dependent frequency of
ion-neutral collisions $\nu _{in}(t)=\left\langle v_{th}\sigma \right\rangle
C_{r}n^{2}t\approx 1\times 10^{-10}n^{2}t_{\mathrm{Kyr}}$ s$^{-1}$. One can
continue to use equations (20)-(21) with the substitution $\nu _{in}=\nu
_{in}(t)$ for the determination of maximum momentum of accelerated
particles. Figure 2 illustrates the results. It is evident that the
ionization of interstellar atoms by supernova radiation diminishes the
collisional wave dissipation.

\section{ Supernova explosion in a stellar wind}

Immediately after the supernova burst, the shock propagates not in the
interstellar medium but through the wind of a progenitor star. The duration of
this phase depends on the wind power and it may cover even the considerable
part of the adiabatic (Sedov) stage of supernova remnant evolution if the
pre-supernova star was very massive, more than about $15$ solar masses. The
massive progenitors of Type-II and Ib supernovae undergo a significant mass
loss prior to explosion during their red and blue giant phases, and the Wolf-Rayet
phase for the most massive stars. The stellar wind modifies the ambient
density distribution and creates an expanding shell. The cosmic ray
acceleration in the case of shock propagation through the wind was considered
by V\"{o}lk and Biermann (\cite{voelkbiermann88}),
Berezinsky and Ptuskin (\cite{berezinsky89}), and Berezhko and
V\"{o}lk (\cite{berezhko00}) under the
assumption of Bohm diffusion for energetic
particles. Bell and Lucek (\cite{bell01}) included
the effect of strong cosmic-ray
streaming instability into the consideration of high velocity shocks.

A distinctive property of the problem is the non-homogeneous distributions of
gas density and magnetic field in the stellar wind flow. In particular, the
spherically symmetric distribution of gas density in the stellar wind is
$n_{w}=$
$\dot M/\left(  4\pi m_{a}u_{w}r^{2}\right)  $, where
$\dot M=10^{-5}\dot M_{-5}$ $M_\odot $ yr$^{-1}$ is the
mass loss, $u_{w}=10^{6}u_{w,6}$ cm s$^{-1}$ is the wind velocity, and the
normalization and numerical values here and below are given for the typical
supernova Type II progenitor, the red giant star
(see Berezinsky and Ptuskin, \cite{berezinsky89}). Similar to
the interplanetary magnetic field, the stellar wind magnetic field has the
shape of a Parker spiral (Parker, \cite{parker58}).
The magnetic field has predominately
azimuthal structure at distance from the star $r\gg u_{w}/(\Omega\sin
\theta)\approx5\times10^{12}$/sin$\theta$ cm where its' value is
$B_{0}=B_{\ast}r_{\ast}^{2}\Omega\sin\theta/(u_{w}r)$. Here $B_{\ast}=10$ G is
the surface magnetic field strength at star radius $r_{\ast}=2\times10^{13}$
cm , $\Omega=2\times10^{-7}$ s$^{-1}$ is the angular velocity of star
rotation, and $\theta$ is the polar angle. Hence $B_{0}(r)r\approx8\times
10^{14}\sin\theta$ G$\times$cm at $r\gg5\times10^{12}$ cm. It is worth
noting that the value of Alfven velocity $V_{a}=B_{0}(r)r\sqrt{u_{w}
/\dot M}$ does not depend on distance $r$ in this case.

The shock propagation in a stellar wind obeys the following
equations (e.g. Lozinskaya \cite{lozinskaya92}):
\begin{eqnarray}
R_{sh}&=&8.77\left(  \frac{u_{w,6}\mathcal{E}_{51}}
{\dot M_{-5}
}\right)  ^{1/3}t_{\mathrm{Kyr}}^{2/3}\mathrm{ pc,}\; \nonumber\\
u_{sh}&=&5.72\times
10^{3}\left(  \frac{u_{w,6}\mathcal{E}_{51}}{\dot M_{-5}
}\right)  ^{1/3}t_{\mathrm{Kyr}}^{-1/3}\;\mathrm{km\ s^{-1}},
\end{eqnarray}
that substitute now equations (1). Expressions (24) are valid at $u_{sh}\gg
u_{w}$, and under the conditions $R_{sh}>R_{0}=M_{ej}u_{w}/\dot M
\approx1\left(  M_{ej}/M_{\odot}\right)  $ pc, $t>t_{0}\approx40\left(
M_{ej}/M_{\odot}\right)  ^{3/2}$ yr when the mass of swept-up gas is
relatively large.

Equations (14), (15) can be used for the estimates of maximum particle
momentum when the instability is balanced by the wave advection and nonlinear
wave dissipation of Kolmogorov type. The asymptotic expression for
$p_{\max}(t)$ at high shock velocity subject to
conditions (16) is the following (compare with equation (17)):
\begin{eqnarray}
\frac{p_{\max}}{mc}&\approx&\frac{2\varkappa aC_{cr}(a)
\xi_{cr}eu_{sh}
^{2}\dot M^{1/2}}{mc^{3}u_{w}^{1/2}}\approx \nonumber\\
&&8.9\times
10^{5}\left(  \xi_{cr}/0.5\right)  \mathcal{E}_{51}^{2/3}\left(  \frac
{u_{w,6}}{\dot M_{-5}}\right)  ^{1/6}t_{\mathrm{Kyr}}^{-2/3}.
\end{eqnarray}

The asymptotic expression at low shock velocity subject to conditions (18) is
\begin{eqnarray}
\frac{p_{\max}}{mc}\approx\frac{0.24\varkappa a^{2}C_{cr}^{2}(a)C_{K}^{3}
\xi_{cr}^{2}\dot M^{1/2}u_{sh}^{7}}{V_{a}^{3}u_{w}^{1/2}c^{5}
}\approx 4.6\times10^{6} \nonumber\\
\left(  \frac{\xi_{cr}}{0.5}\right)  ^{2}
\mathcal{E}_{51}^{7/3}\left(  \frac{u_{w,6}}{\dot M_{-5}
}\right)  ^{1/3}\left(  \frac{B_{\ast}r_{\ast}}{8\times10^{14}}\right)
^{3}t_{\mathrm{Kyr}}^{-7/3},
\end{eqnarray}
at $\theta\approx\pi/2$ (compare with equation (19)).

In actuality the regime (26) can hardly be achieved since the size of the wind
of a red giant star usually does not exceed $10$ pc and the supernova shock
breaks out of the wind cavity at $t$ $\lesssim1$ Kyr when the shock velocity
is still high. Also, one has to keep in mind that the wind parameters are
significantly changing over the course of star evolution.

Equations (25), (26) show that, at the accepted
values of parameters, the supernova shock moving in the wind
of a red supergiant star can accelerate cosmic rays up to somewhat higher
energies and the decrease of $p_{\max}$ with time is weaker than in the
case of a shock moving in homogeneous interstellar medium. Also, at a given
age of supernova remnant, the value of
$p_{\max}$ is essentially higher for the explosion in wind.

\section{Discussion and conclusion}

The generation of turbulence by cosmic-ray streaming instability ahead of the
shock is an integral part of cosmic ray acceleration in supernova remnants.
We considered this process without limitation on the amplitude of random
magnetic field and with allowance for the wave dissipation through Kolmogorov
type nonlinearity and by ion-neutral collisions. The amplification of random
magnetic field at high shock velocity, $u_{sh}\gg10^{3}$ km s$^{-1}$, leads to the
increase of maximum energy of accelerated particles above the Bohm limit
calculated at the interstellar magnetic field strength. This is in agreement
with the results of Bell \& Lucek (\cite{bell01}).
On the other hand, in support and
development of earlier investigations of V\"{o}lk et al.
(\cite{voelk88}), Fedorenko
(\cite{fedorenko90}), and Drury et al. (\cite{drury96})
who dealt with not very strong turbulence and
considered different nonlinear processes,
we found that the nonlinear and
linear wave damping may considerably suppress the level of turbulence at $%
u_{sh}\lesssim10^{3}$ km s$^{-1}$ and thus may decrease the maximum particle energy
much below the Bohm limit. On the whole, the maximum energy of accelerated
particles with charge $Ze$ rapidly decreases from $\gtrsim$ $10^{14}Z$ eV to
much smaller energies $\sim10^{10}Z$ eV during the Sedov stage of supernova
remnant evolution at the age from a few hundred years to about $10^{5}$ yr.
Our main results are illustrated by Figures 1 and 2 for the cases of supernova
bursts with kinetic energy of ejecta $\mathcal{E}=10^{51}$ erg in hot
rarefied interstellar gas (the gas temperature $T=7\times10^{5}$ K, the
average gas density including clouds $n_{0}=10^{-2}$ cm$^{-3}$, the density
of diffuse gas $n=3\times 10^{-3}$ cm$^{-3}$, the interstellar magnetic
field $B_{0}=2$ $\mu$G) and in the warm weakly ionized interstellar gas ($%
T=8\times10^{3}$ K, $n_{0}=0.4$ cm$^{-3}$, $n=0.1$ cm$^{-3}$, $B_{0}=5$ $\mu$%
G) respectively. The assumed efficiency of cosmic ray acceleration is $%
\xi_{cr}=0.5$.

It is remarkable that the highest energy $E_{\max }$ which the cosmic ray particle can
reach in the process of acceleration does not explicitly depend on the value
of interstellar magnetic field, see equations (17), (21). It is easy to
check that the highest energy is attained at the time of transition
from the free expansion to Sedov stage of supernova remnant evolution. Using
equation (19), one can make the following estimate at $t=t_{0}$ (Sedov
solution is not applicable yet at this moment):

\begin{eqnarray}
E_{\max} &\approx &\frac{2\varkappa aC_{cr}(a)\xi
_{cr}Zmcu_{0}^{2}R_{0}}{V_{a}r_{g0}}\approx \nonumber \\
&&8\times 10^{16}Z\left( u_{0}/3\times 10^{4}\mathrm{ km\ s^{-1}}\right)^{2}
M_{ej}^{1/3}n^{1/6}
\mathrm{ eV}
\end{eqnarray}
at $\xi _{cr}=0.5$, $\varkappa =0.1$, $a=0.3$, $n_{0}=n$,
the mass of ejecta $M_{ej}$ is in solar masses. The estimate (27) is in
agreement with the estimates made by Bell \& Lucek (2001). The corresponding
random magnetic field comprises
\begin{eqnarray}
\delta B_{\max } &\approx &\frac{2aC_{cr}(a)\xi _{cr}u_{0}}{3V_{a}}
B_{0}\approx  \nonumber \\
&&3\times 10^{-3}\xi _{cr}\left( u_{0}/3\times 10^{4}
\mathrm{ km\ s^{-1}}\right) n^{1/2}
\mathrm{ G}
\end{eqnarray}
in the upstream region of the size $D(p_{\max })/u_{0}\approx \varkappa
R_{0}\approx 0.1R_{0}$. The field can reach more than $\sim 10^{-3}$ G in the
downstream compressed region behind the shock. The value of $E_{\max }$
given by equation (27) is by a factor of $\delta B_{\max }/B_{0}\sim 100$
larger than the Bohm value calculated for the interstellar field $B_{0}$.

Equation (25) refers to the case of supernova explosion in the wind of progenitor
star with the characteristic mass loss rate $\dot M=10^{-5}$
 $M_\odot $ yr$^{-1}$ and
the wind velocity $u_{w}=10^{6}u_{w,6}$ cm s$^{-1}$.

The estimate of $E_{\max}$ in the case of supernova explosion in the wind of a
progenitor star is:
\begin{eqnarray}
E_{\max} \approx \frac{2\varkappa aC_{cr}(a)\xi_{cr}Zeu_{sh}^{2}
\dot M^{1/2}}{cu_{w}^{1/2}}\approx \nonumber \\
7\times10^{16}Z\left(  u_{0}
/3\times10^{4}\mathrm{km\ s^{-1}}\right) ^{2}
\sqrt{\dot M_{-5}/u_{w,6}}\mathrm{ eV},
\end{eqnarray}
that in fact coincides with the estimate (27). In both cases the highest
particle energy can hardly exceed $10^{17}Z$ eV even for the extreme value of
shock velocity $u_{0}=3\times10^{4}$ km s$^{-1}$.

Two essential assumptions are crucial for our conclusions. The first is the
high efficiency of shock acceleration that transfers the fraction $\xi
_{cr}\approx0.5$ of the upstream momentum flux entering the shock to the
cosmic ray pressure and strongly modifies the shock profile. This assumption
is based on the results of numerical simulations of diffusive shock
acceleration by Berezhko et al.
(\cite{berezhko96}), Berezhko \& Ellison (\cite{berezhko99}). The
second essential assumption is the high rate of nonlinear Kolmogorov-type
dissipation of magnetohydrodynamic turbulence that balance the cosmic ray
streaming instability. This assumption is supported by the analytical and numerical
investigations of MHD turbulence, see e.g.
Goldreich \& Sridhar (\cite{goldreich95}),
Verma et al. (\cite{verma96}), and by the
observations of Kolmogorov type spectra of
turbulence in the interplanetary (Burlaga, \cite{burlaga95})
and interstellar (Armstrong
et al., \cite{armstrong95}) space.
It should be pointed out that our description of
nonlinear wave interactions is very simplified and can be considered as not
more than an order of magnitude estimate. In particular, we ignore the
complicated anisotropic structure of magnetohydrodynamic turbulence in
$\mathbf{k}$-space.

In the case of supernova remnant expansion into partly
ionized interstellar gas, the wave dissipation by ion-neutral collisions was
also taken into account. Collisional damping may suppresses particle
acceleration more than the nonlinear wave dissipation does if the
surrounding interstellar gas is not vastly ionized by the supernova burst
(the case of Type Ia supernova).

The fast decrease of maximum energy of cosmic rays in supernova remnants
with time may help to explain why the shell remnants at the age more than a
few thousand years are not prominent sources of TeV gamma rays produced by
very high energy cosmic rays (see e.g. Buckley et al.
\cite{buckley98}) in spite of the
optimistic preliminary theoretical estimates
(Drury et al. \cite{drury94}, Naito \&
Takahara \cite{naito94}) and in spite of a
few detections of gamma-rays from these
sources at about $100$ MeV (Esposito et al. \cite{esposito96}).
On the other hand, a
number of young supernova remnants (SN1006, Cas-A, Tycho, Kepler,
RXJ1713.7-3946) were reported as cosmic ray accelerators of very high energy
particles with energies up to $\gtrsim10$ TeV based on the observations of
nonthermal synchrotron X-rays or/and TeV gamma rays. (The effect for the old
remnants is evidently compounded by the specific selection of the objects
near massive gas clouds that increases the generation of secondary gamma
rays through $\pi^{0}$ decays whereas the high gas density $n_{0}$ reduces
$p_{\max}$ as it is clear from equations (19), (22).) The variety of
theoretical constructions by other authors were suggested to resolve this
problem (Gaisser et al. \cite{gaisser98}, Baring et al.
\cite{baring99}, Kirk et al. \cite{kirk01}, V\"{o}lk
\cite{voelk01}, Malkov et al. \cite{malkov02})
but the ultimate choice will be probably made when
the data from a new generation of ground-based (CANGAROO-III, HESS, VERITAS,
MAGIC) and space (GLAST) gamma-ray experiments in conjunction with new X-ray
satellites will be available.

The accounting for non-linear effects in the instability that accompanies the
acceleration of cosmic rays in supernova remnants may simultaneously
eliminate two difficulties of the modern cosmic ray astrophysics. It raises
the maximum energy of accelerated particles in young supernova remnants
above the standard Bohm limit through the production of strong random
magnetic fields and thus helps to explain the origin of galactic cosmic rays
with energies up to $\sim10^{17}Z$ eV, though it does not cover energies
larger than about $3\times10^{18}$ eV even for Iron nuclei and this implies the
extragalactic origin for cosmic rays with the highest observed energies. It also
decreases the maximum energy of particles in late Sedov stage of supernova
remnant evolution that allows to explain why these objects are not bright in
very high energy gamma rays.

\begin{acknowledgements}

This work was stimulated by the discussions with F. Aharonyan, E. Berezhko, L.
Ksenofontov, and H. V\"{o}lk during our teamwork on the project ''Energetic
particles in the Galaxy: acceleration, transport and gamma-ray production'' at
the International Space Science Institute (Bern). The authors are grateful to
Johannes Geiss for sponsorship and kind hospitality at ISSI. This work was
supported by RFBR-01-02-17460 grant at IZMIRAN. VSP was also supported by the
NASA grant NAG5-11091 during his visit to the University of Maryland where
part of this work was fulfilled.
\end{acknowledgements}

\end{document}